\documentstyle[12pt,blois]{article}

\def\msun{{\rm\,M_\odot}}

\begin{document}
\heading{THE GALAXY HALO FORMATION RATE}
\author{W. Percival \& L. Miller} 
{Dept. of Physics, University of Oxford, Oxford OX1 3RH, U.K.} 

\begin{bloisabstract}
The rate at which galaxy halos form is thought to play a key role in
explaining many observable cosmological phenomena such as the initial
epoch at which luminous matter forms and the distribution of active
galaxies. Here we show how Press-Schechter theory$^{\cite{ps}}$ can be
used to provide a simple, completely analytic model of the halo
formation rate. This model shows good agreement with both Monte-Carlo
and N-body simulation results.
\end{bloisabstract}

\section{Introduction to Press Schechter Theory} \label{sec:ps}
Following standard Press-Schechter (PS) theory, we convolve a
homogeneous, isotropic Gaussian random field representing primordial
density fluctuations with a sharp k-space filter. The filter size is
related to mass of halo by $M=3/4\pi^{2}\rho R^{3}$ $^{\cite{lc93}}$.
The overdensity of the filtered field at any point traces a
Brownian random walk with the variance of the filtered field being the
`time' axis$^{\cite{bond,peacock}}$. Regions are considered to have
collapsed to form dark halos of mass $>M$ if this overdensity exceeds a
critical overdensity, $\delta_c(t)$. The distribution of halo masses
is given by the distribution of upcrossings of the line
$\delta=\delta_{c}(t)$, a solution of the diffusion equation
with absorbing boundary condition$^{\cite{bond}}$. This leads to a
formula for the comoving number density of objects of mass $M$ given
time t, $n(M,t)$, the standard PS result :
\begin{equation}
  n(M,t)MdM=\rho P(M|\delta_{c}(t))dM
  =2\rho\frac{\delta_{c}(t)}{(2\pi)^{1/2}\sigma_{M}^{2}}\exp\left(
  -\frac{\delta_{c}(t)^{2}}{2\sigma_{M}^{2}}\right)\left|\frac{d\sigma_{M}}{dM}
  \right|dM.    \label{eq:ps}
\end{equation}
We normalise the power spectrum by setting the variance of the
density field, filtered with a top-hat filter of radius
$8h^{-1}$\,Mpc, $\sigma_8=0.64$. The critical density used with the
top-hat filter is that predicted from uniform spherical
collapse$^{\cite{peebles}}$, that used with a sharp k-space filter is
set so that we predict the same number density of halos with both
filters at the normalisation mass.

\section{The Halo Formation Rate}
\begin{figure}[ht]
  \centering
  \begin{picture}(0,0)%
  \includegraphics{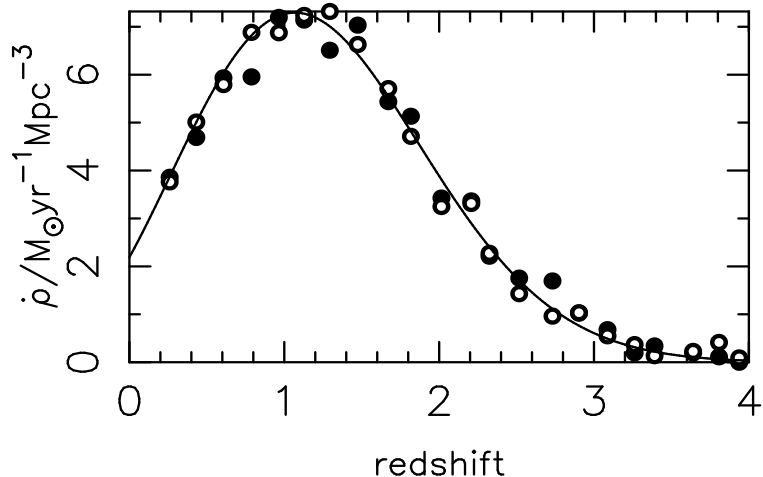}%
  \end{picture}%
  \setlength{\unitlength}{0.00083300in}%
  \begingroup\makeatletter\ifx\SetFigFont\undefined
  % extract first six characters in \fmtname
  \def\x#1#2#3#4#5#6#7\relax{\def\x{#1#2#3#4#5#6}}%
  \expandafter\x\fmtname xxxxxx\relax \def\y{splain}%
  \ifx\x\y   % LaTeX or SliTeX?
  \gdef\SetFigFont#1#2#3{%
    \ifnum #1<17\tiny\else \ifnum #1<20\small\else
    \ifnum #1<24\normalsize\else \ifnum #1<29\large\else
    \ifnum #1<34\Large\else \ifnum #1<41\LARGE\else
       \huge\fi\fi\fi\fi\fi\fi
    \csname #3\endcsname}%
  \else
  \gdef\SetFigFont#1#2#3{\begingroup
    \count@#1\relax \ifnum 25<\count@\count@25\fi
    \def\x{\endgroup\@setsize\SetFigFont{#2pt}}%
    \expandafter\x
      \csname \romannumeral\the\count@ pt\expandafter\endcsname
      \csname @\romannumeral\the\count@ pt\endcsname
    \csname #3\endcsname}%
  \fi
  \fi\endgroup
  \begin{picture}(4790,3057)(1114,-4156)
  \end{picture}
  \caption{
  Plot comparing Monte-Carlo (open symbols) and N-body (solid symbols)
  results with PS predictions (solid line) of the 
  formation rate of halos of mass $1.3\times10^{13}\msun$.
  The Monte-Carlo analysis involved
  $10^{4}$ halo mergers recorded in $\sim10^{6}$ walks, each
  consisting of $2^{11}$ uniform steps in $\sigma_{M}^{2}$.  
  The Hydra N-body, hydrodynamics code$^{\cite{couchman}}$ was used to
  run a simulation with $128^{3}$ dark matter particles in a flat
  $\Lambda=0$ universe with $h=0.5$ and standard CDM power
  spectrum with shape parameter $\Gamma=0.25$. Groups of particles were found
  using a standard friends of friends (FOF) algorithm with
  linking length $b=0.2$.} 
\label{fig1}
\end{figure}

We now wish to calculate the distribution of cosmic epochs at which
halos of a given mass $M$ are formed, $P(t|M)$. Because of the nature 
of the random walks if we make $\delta_{c}$ bound,
$\delta_{min}\leq\delta_{c}\leq\delta_{max}$, it can be thought of as
a random variable with uniform probability density. We can remove
these bounds later without affecting the result. Using Bayes'
theorem and $P(\sigma^{2}|\delta_{c})$, calculated in the previous
section, we obtain the conditional probability
$P(\delta_{c}|\sigma_{M}^{2})$. Taking the limit as
$\delta_{min}\rightarrow0$ and $\delta_{max}\rightarrow\infty$ and
changing the variables to mass $M$ and time $t$, we obtain the
probability that a halo of mass $M$ formed in the time interval ($t$,$t+dt$) :
\begin{equation}
  P(t|M)dt=\frac{\delta_{c}}{\sigma_{M}^{2}}
    \exp\left(-\frac{\delta_{c}^{2}}{2\sigma_{M}^{2}}\right)
    \left|\frac{d\delta_{c}}{dt}\right|dt. \label{eq:rate_cstm}
\end{equation}
The predictions of equation~\ref{eq:rate_cstm} are compared with
the results of both Monte-Carlo analysis and an N-body simulation in
Fig.~\ref{fig1}. Excellent agreement is found between all results. 
\section{Conclusions}
We have shown that we can use Press Schechter theory to produce a
simple analytic model of the formation rate of dark halos. Our model uses
statistical analysis of the trajectories and overcomes the problem of
relating a trajectory through the hierarchical build up of
structure to a particular halo. We have shown that the results of
this model agree well with those from both Monte-Carlo analysis
and from N-body simulations. We are grateful for the use of the Hydra
N-body code \cite{couchman} kindly provided by the Hydra consortium.

% References listed in alphabetical order ...

%
\vfill
\end{document}